\documentclass{article}
\usepackage[utf8]{inputenc}
\usepackage{authblk}
\usepackage{hyperref}
\usepackage{bm}

\title {A comment on: Adam R. Brown $\&$ Leonard Susskind’s paper “A holographic wormhole traversed in a quantum computer”}
\author{Galina Weinstein}
\affil{\normalsize Reichman University, The Efi Arazi School of Computer Science, Herzliya; University of Haifa, The Department of Philosophy, Haifa, Israel.} 

\begin{document}

\maketitle

\begin{abstract} 

Adam Brown and Leonard Susskind write in their new paper, “A holographic wormhole traversed in a quantum computer” (\cite{Brown}): “The idea of a wormhole dates back to 1935, when Albert Einstein and his collaborator, Nathan Rosen, studied black holes in the context of Einstein’s general theory of relativity. […]  In the same year, Einstein and Rosen wrote another paper, this time in collaboration with Boris Podolsky. […]  At the time, these two ideas — wormholes and entanglement — were considered to be entirely separate”. I will show in this comment that for Einstein, the Einstein-Rosen (ER) bridge and the EPR argument were not "entirely separate". 
      
\end{abstract}

\section{Introduction}

Adam Brown and Leonard Susskind write in their new paper, “A holographic wormhole traversed in a quantum computer” ("A holographic wormhole
in a quantum computer") (\cite{Brown}): 

\begin{quote}
“The idea of a wormhole dates back to 1935, when Albert Einstein and his collaborator, Nathan Rosen, studied black holes in the context of Einstein’s general theory of relativity. […]  In the same year, Einstein and Rosen wrote another paper, this time in collaboration with Boris Podolsky. The trio’s paper
examined quantum mechanics (without gravity), and identified the phenomenon now known as quantum entanglement, which Einstein described as 'spooky action
at a distance'.[…]  At the time, these two ideas — wormholes and entanglement — were considered to be entirely separate”.     
\end{quote}

Brown and Susskind's description uses a modern point of view which is very different from Einstein's own perception and motivation. For Einstein in 1935, the ER bridge was not a wormhole, i.e., a topological shortcut through spacetime. Neither was it two black holes that are taken to be the mouths of a wormhole.  Einstein did not even speak of “black holes”. The above modern interpretation of the ER bridge probably goes back to Robert Fuller and John Archibald Wheeler (\cite{Fuller}), and is found in the more recent paper by Juan Maldacena and Susskind, "Cool horizons for entangled black holes" \cite{Maldacena2}:

\begin{quote}
"General relativity contains solutions in which two distant black holes are connected through the interior via a wormhole, or Einstein-Rosen bridge".     
\end{quote}

In this comment, I want to show that the Einstein-Rosen (ER) bridge and the Eisnstein-Podolsky-Rosen (EPR) argument were not ”entirely separate” to use Brown and Susskind's words and they were both related to Einstein's work on unified field theory.

\section{The EPR thought experiment} \label{2}

In early March 1935 Einstein, Rosen, and Podolsky submitted the manuscript of the EPR paper, "Can Quantum-Mechanical Description of Physical Reality Be Considered Complete?", to \emph{Physical Review}, and it was published on May 15, 1935 (\cite{Einstein2}).

Einstein, Rosen, and Podolsky used the EPR thought experiment to demonstrate that the Heisenberg uncertainty principle fails. In 1927, Heisenberg wrote the following relation for two operators $P$ and $Q$ corresponding to momentum and position that do not commute:

\begin{equation} \label{equation1}
PQ-QP = \frac{h}{2\pi i} 
\end{equation}

\noindent where $h$ denotes Planck’s constant. 

In their paper, Einstein, Rosen, and Podolsky set themselves to challenge equation \ref{equation1}.
According to Einstein, Rosen, and Podolsky, since at the time of measurement the two particles no longer interact, no change can take place in the second particle $B$ as a result of measurements performed on particle $A$.
The momentum of particle $A$ is measured and it is found to have a value of $p$. According to the conservation of linear momentum, the momenta are equal and opposite. Thus, without having disturbed the remote particle $B$ we know that its momentum is $-p$.

$\psi_p(x_2)$ is an eigenfunction of the operator $P$:

\begin{equation} \label{equation2}
P = \frac{h}{2\pi i} \frac{\partial \psi_p (x_2)}{ \partial (x_2)}
\end{equation}

\noindent corresponding to the eigenvalue $-p$ of the momentum of particle $B$. Therefore, $P$ is an “element of reality” because we have not disturbed particle $B$.

The same reasoning applies to the position of particle $B$. According to “conservation of relative position”, if we measure the position of $A$ along the $x$-axis to be $x_1$, we know the position of $B$, which is $x_2 = x_1 + x_0$, where $x_0$ is the separation between the two particles. A measurement of the position of $A$ allows one equally well to predict with certainty the position of a remote particle $B$, again without in any way disturbing $B$.
$\phi_x(x_2)$ is an eigenfunction of the operator $Q$:

\begin{equation} \label{equation3}
Q = (x_2)
\end{equation}

\noindent corresponding to the eigenvalue $x_2$ of the position of $B$. Therefore, $Q$ is an “element of reality.”

Equations (\ref{equation2}) and (\ref{equation3}) contradict equation (\ref{equation1}). Thus, “it is in general possible for $\psi_p(x_2)$ and $\phi_x(x_2)$ to be eigenfunctions of two non-commuting operators [$P$ and $Q$], corresponding to physical quantities,” momentum and position of particle $B$.
By way of contrast, “[i]n quantum mechanics it is usually assumed” that either the momentum or the position of $B$ “correspond exactly to what can be measured” on particle $A$.

Einstein, Rosen, and Podolsky end their paper by saying that they are thus forced to conclude that the quantum mechanical description of physical reality given by wavefunctions is not complete (\cite{Einstein2}).

It was Podolsky who wrote the EPR paper and Einstein was unhappy with the way it turned out. Einstein believed that the theoretical physicist starts from free creations of the human mind, i.e., imagination. These are independent of any experiment. The physicist creates a heuristic theory of principle. For instance, Einstein created special relativity and formulated the two principles of the theory. Once a principle theory is created, the physicist trusts their intuition and inspiration and they are not surprised when the experiment confirms their intuitions. Einstein was, therefore, guided by principles. 

In 1935, Einstein presented a few principles. One can formulate these principles as follows:

1) \emph{Causality}: Suppose that $A$ and $B$ are spacelike separated (no signal can travel from $A$ to $B$, unless faster than the speed of light). Particle $A$ cannot instantaneously affect particle $B$.

2) \emph{Locality}: At the time of measurement, the two particles no longer interact. No real change can take place in $B$, being far away from $A$, as a result of measurements performed on $A$.

3) \emph{Separability}: The real state of $B$ cannot depend upon the kind of measurement we carry out on $A$. That is, two systems separated in space have their own separate real states.

In 1948, Einstein combined the first two principles, locality-causality with the third, separability, and formulated one single principle:

4) \emph{Contiguity} which requires no spooky action-at-a-distance between particles $A$ and $B$ and demands a timelike signaling which would travel from $A$ to $B$ with a velocity smaller than the speed of light.
This principle is satisfied in any field theory such as the general theory of relativity and unified field theories.
\vspace{1mm} 

Let us see how these principles guided Einstein in his efforts at trying to solve the EPR paradox. 
After the publication of the EPR paper, Einstein told Erwin Schrödinger that it was Podolsky who wrote the EPR paper. Einstein added that he was unhappy with the way it turned out, and explained his version of the EPR argument: consider two quantum mechanical systems, $A$ and $B$ that interact and then separate (Einstein to Schrödinger, June 19, 1935, \cite{Howard}): 

\begin{quote}
"After the interaction, the real state of $AB$ consists precisely of the real state of $A$ [$\psi_A$] and the real state of $B$ [$\psi_B$], which two states have nothing to do with one another. The real state of $B$ thus cannot depend upon the kind of measurement I carry out on $A$ ('Separation Hypothesis'…). But then, for the same state of $B$, there are two (in general arbitrarily many) equally justified [$\psi_B$], which contradicts the hypothesis of a one-to-one or complete description of the real states".     
\end{quote}

In 1948 Einstein wrote (A. Einstein to M. Born, April 5, 1948, Letter 88, \cite{Born}): 

\begin{quote}
"The following idea characterizes the relative independence of objects far apart in space ($A$ and $B$): external influence on $A$ has no direct influence on $B$; this is known as the ‘principle of contiguity' which is used consistently only in the field theory.

I now make the assertion that the interpretation of quantum mechanics [. . . ] is not consistent with principle II [the principle of contiguity]. Let us consider a physical system $S_{12}$, which consists of two part-systems $S_1$ and $S_2$ [. . . they interact and are described by $\psi_{12}$ . . . ]. At time $t$ let the two part-systems be separated from each other in space [. . . ].
\end{quote}

\noindent And Einstein continues to describe the EPR thought experiment. 

From Einstein’s point of view, the EPR paradox could be solved by either “spooky action at a distance” or by violating the “Separation Hypothesis” (separability) and the principle of contiguity. Einstein could certainly not accept “spooky action at a distance”. That is because he spoke about the “real state of $B$”. As I show in my paper (\cite{Weinstein}), he could not violate separability and the principle of contiguity either because he treated the EPR pair of particles as a two-body problem in the field theory. And so, Einstein stubbornly insisted on the separability of the two particles, even at the cost of complicating the explanation of the EPR thought experiment to the point of creating a paradox. 

\section{The ER bridge paper} \label{1}

A few days before the EPR paper was published, Einstein and his assistant Rosen submitted the manuscript of the ER bridge paper, "The Particle Problem in the General Theory of Relativity", to \emph{Physical Review}, and it was published on July 1, 1935 (\cite{Einstein1}).

In the introduction to their paper, Einstein and Rosen write (\cite{Einstein1}):

\begin{quote}
“As is well known, Levi-Civita and Weyl have given a general method for finding axially symmetric static solutions of the gravitational equations. By this method one can readily obtain a [Ludwig Silberstein's static two-body] solution which, except for two point singularities lying on the axis of symmetry, is everywhere regular and is Euclidean at infinity. Hence, if one admitted singularities as representing particles one would have here a case of two particles not accelerated by their gravitational interaction, which would certainly be excluded physically".     
\end{quote}

As I show in my paper (\cite{Weinstein}), Einstein thought he had found a solution to the above problem, in the form of an ER bridge. He tried to exclude singularities from the theory, and at the same time, elementary material particles would not have to be represented as singularities of the field. 

Einstein and Rosen first modified the field equations in order to obtain a "solution free from singularities" (a singularity-free solution): particles are no longer represented by singularities in the field but they are rather modeled by a topological structure, a bridge. Hence, Einstein tried to replace an elementary particle having mass but no charge with a topological structure, an ER bridge of finite length in four-dimensional spacetime. The spatially finite bridge was identified as being the neutron, which is an elementary particle having mass but no electric charge. With this conception, Einstein thought he could represent an elementary particle using only the field equations and not as singularities in the field. Einstein and Rosen then tried to replace an elementary charged particle with a spatially finite bridge. In order to obtain a bridge that would represent a charged particle, they took into consideration the negative of the Maxwell stress-energy tensor in vacuum. Einstein and Rosen set the mass equal to zero and obtained a solution that is free from singularities for all finite points in the space of two sheets and the charged particle was again represented by a bridge between the two sheets. The ER bridge represented an elementary electric particle with electric charge but without mass and with negative energy density. 
Einstein and Rosen obtained two types of bridges connecting the two identical flat sheets. Each type of particle, having either mass or electric charge, is represented by a different type of bridge connecting the two flat sheets. 

Einstein desired “to provide a unified foundation on which the theoretical treatment of all phenomena could be based". He firmly believed that "the method of general relativity provides the possibility of accounting for atomic phenomena” (\cite{Einstein1}). Einstein’s unified field theory was supposed to account for the existence of two elementary particles: the electron and the proton (\cite{Sauer}). Hence, the electron and proton would each be represented by two bridges between the two congruent identical flat sheets. Einstein and Rosen expected that “processes in which several elementary particles take part correspond to regular solutions of the field equations with several bridges between the two equivalent sheets corresponding to the physical space”. 
That is, Einstein and Rosen tried to find a solution to the modified field equations, "free from singularities" (a singularity-free solution) for several particles, in which several bridges are connected to the two identical flat sheets. However, the major trouble was: “For the present one cannot even know whether regular solutions with more than one bridge exist at all” (\cite{Einstein1}). 

As I show in my paper \cite{Weinstein}, Podolsky worked with Einstein on the multi-bridge (many-body) problem in the ER bridge theory. This is evident from a letter that Einstein sent to his closest friend Michele Besso in early 1936 describing to him his work on the ER bridge (Einstein to Besso, February 16, 1936, EA 7-372.1 in \cite{Besso}):

\begin{quote}
    
"Enclosed I am sending you a short paper, which represents the first step. The neutral and the electric particles appear, so to speak, as a hole in space, in such a way that the metric field returns into itself. Space is described as double sheets. In Schwarzschild's exact spherical symmetric solution, the particle appears in ordinary space as a singularity of the type $1-2m/r$. Substituting $1 – 2m = u^2$, the field becomes regular in $r-$ space. When $u$ extends from $-\infty$ to $+\infty$, $r$ extends from $+\infty$ to $r = 2m$ and then back to $r = +\infty$. This represents both 'sheets' in Riemann's sense, which are joined by a 'bridge' at $r = 2m$ or $u = 0$. It is similar to the electric charge.
A young colleague (Russian Jew) and I are relentlessly struggling with the treatment of the many-body problem on that basis. But we have already overcome the serious difficulties of the problem, so it will soon become clear what the matter is". 

\end{quote}

The “Russian Jew” is Boris Podolsky. 
\vspace{1mm} 

\section{Unified Field theory} \label{3}

In the case of the ER bridge, Einstein was inspired by Silberstein’s static two-body solution (problem) in the field theory to replace an elementary particle with a topological structure, an ER bridge. The ER bridge served as a model for the particle. In the case of the EPR argument, Einstein linked the separability of $A$ and $B$ with the field theory and treated the two particles $A$ and $B$ as if they were two particles in the field theory. 

One could argue that Einstein’s work on ER and EPR boils down to the argument that, the ER bridge and the EPR argument represent different ways of interpreting the two-body problem in the field theory. That is because historically, "the Einstein-Rosen bridge paper was part of an ongoing controversy with Ludwig Silberstein" over the two-body problem in the field theory (\cite{Kennefick}). And in 1935 and afterward, the non-separability of $A$ and $B$ was Einstein's most fundamental problem with quantum mechanics.

But I don’t agree with this point, because I argue that this way of expressing Einstein’s work on the ER bridge theory and the EPR argument can be misleading. Remember that Einstein and Rosen had long tried to find a multi-bridge solution, i. e., two or more bridges between the two congruent identical flat sheets would serve as a model for several elementary particles.  

I suggest that a better approach to Einstein’s work during 1935-1936 is to say that, what made the ER bridge connected with the EPR thought experiment depended on Einstein’s worldview. This is sometimes expressed as the question of whether one can exclude singularities from the field theory, as separability and contiguity. I further elaborate on this point now.

Maldacena writes in his paper, "Black Holes, Wormholes and the Secretes of Quantum Spacetime" (\cite{Maldacena}): 

\begin{quote}

"Interestingly both quantum entanglement and wormholes date back to two articles written by Albert Einstein and his collaborators in 1935. On the surface, the papers seem to deal with very different phenomena, and Einstein probably never suspected that there could be a connection between them".    
\end{quote}

And Maldacena wrote in 2013, in an article on the website of the \emph{Institute for Advanced Study}, "Entanglement and the Geometry of Spacetime":\footnote{https://www.ias.edu/ideas/2013/maldacena-entanglement} 
\begin{quote}
"In 1935, Albert Einstein and collaborators wrote two papers at the Institute for Advanced Study. One was on quantum mechanics [1] and the other was on black holes [2]. The paper on quantum mechanics is very famous and influential. It pointed out a feature of quantum mechanics that deeply troubled Einstein. The paper on black holes pointed out an interesting aspect of a black hole solution with no matter, where the solution looks like a wormhole connecting regions of spacetime that are far away. Though these papers seemed to be on two completely disconnected subjects, recent research has suggested that they are closely connected".    
\end{quote}

Like Brown and Susskind in their paper (\cite{Brown}), Maldacena attributes to Einstein views that reflect modern thinking which is very different from Einstein's original thinking and motivation.

Einstein did not speak of wormholes and obviously, ER bridges differ from wormholes. The very idea of a wormhole was not even invented in Einstein’s lifetime. Wheeler invented wormholes in 1957. In his paper, “The Present Position of Classical Relativity Theory and Some of its Problems”, presented in 1957 (after Einstein’s death) at a conference in Chapel Hill, North Carolina, Wheeler included a drawing of a two-mouth wormhole. The two mouths $P_1$ and $P_2$ are connected to the same “upper region”, i.e., spacetime $A$. 
Wheeler then writes (\cite{Wheeler}): 

\begin{quote}
“If one thinks of the upper region $A$ as a two-dimensional space, there exists the possibility of connecting two regions $P_1$ and $P_2$ by a ‘wormhole’ so that for instance an ant coming to $P_1$, would emerge at $P_2$ without ever having left the two-dimensional surface on which he started out”.     
\end{quote}

It is neither physically satisfying nor historically accurate to say that the above wormhole is an ER bridge (though it is reasonable that Wheeler was inspired by Einstein's topological structure of 1935). Similarly, the multi-bridge problem (presented in the previous section \ref{2}) obviously has no relationship to modern suggestions. e.g., multi-mouth wormholes (a wormhole that has more than two mouths).

If there is one thing I know for certain, it is that for Einstein in 1935, the ER bridge was not a wormhole, i.e., a topological shortcut through spacetime.  

Einstein's ER bridge was intimately linked with the "Schwarzschild Singularity". I will try to explain how both are connected. Schwarzschild showed that as follows from his equations, a collapsing sphere (a star) of a gravitational mass cannot have a radius, measured from outside, whose numerical value is less than $r = 2GM$. If it does have a radius $r > 2GM$, then the equations break down, i.e., the sphere collapses (\cite{Schwarzschild}). 

In 1935, the radius $r = 2GM$ was called the "Schwarzschild Singularity"; it limits the size of the sphere to the size $r > 2GM$. Einstein thought that it was meaningless to speak of what occurs beyond the "Schwarzschild Singularity" because what occurs beyond this area has no physical meaning. 
As Einstein understood it, when we speak about the "Schwarzschild Singularity", we mean $r = 2GM$, and not $r = 0$. Therefore, obviously, Einstein did not speak of "black holes" either. 

In 1970, after Wheeler had named the region of space-time beyond the Schwarzschild singularity a "black hole", Rosen, as if still under the spell of his great deceased mentor Einstein, was still somewhat skeptical about the physical reality of the area beyond the Schwarzschild singularity. In a relativity conference in the Midwest, United States, he said that the portion of space corresponding to $r > 2GM$ is "non-physical". He thought that even a coordinate transformation that removes the singularity could not change this situation. What it means is that the surface $r = 2GM$ represents the boundary of physical space and should be regarded as an impenetrable barrier for particles and light rays (\cite{Rosen}). 

As Einstein told Besso in 1936, Einstein and Rosen performed a coordinate transformation to remove the region containing the Schwarzschild singularity $r = 2GM$. They introduced in place of the Schwarzschild singularity a new variable and the Schwarzschild solution became a regular solution, free from singularities for all finite points and for all values of the new variable. The new solution was a mathematical representation of physical space by a space of two congruent identical flat sheets that were joined by a bridge at $r = 2GM$. 
\vspace{1mm} 

I wish to emphasize that during 1935-1936, Einstein was working on his unified field theory. He wrote in 1936 in his paper, "Physics and Reality" (\cite{Einstein3}):

\begin{quote}

"0nly the examination of the many-bridge-problem can show whether 
or not this theoretical method [...] reproduces the facts which quantum mechanics has so wonderfully comprehended".     
\end{quote}

He made "Many fruitless efforts to find a field representation of matter free from singularities based" on a unified field theory program. And, as he wrote in 1938 with Peter Bergmann (\cite{Einstein1}):

\begin{quote}
"We tried to find a rigorous solution of the gravitational equations, free from singularities, by taking into account the electromagnetic field. [...] Our investigation was based on the theory of 'bridges' (Einstein and Rosen, Phys. Rev., 48: 73 (1935). We convinced ourselves, however, that no solution of this character exists". 
\end{quote}

Einstein believed that quantum mechanics could not serve as a new theoretical basis for physics because it was an incomplete representation of real physical systems. The EPR paper showed this through the EPR argument. But Einstein, being guided by separability (and later by contiguity), strove to derive the results of quantum mechanics from a classical unified field theory program. Jeroen van Dongen writes in his book, \emph{Einstein's Unification} that, this was "Einstein’s larger research program, with its ultimate goal of rederiving from classical field theories the quantum nature of matter". But before taking up this issue, "Einstein had studied what general relativity might have to offer in this respect. Together with Rosen he formulated what is now generally referred to as the Einstein–Rosen bridge, originally proposed as an attempt to unify field and particle concepts" (\cite{Dongen}).  

It is our modern point of view (since Stephen Hawking’s studies in the 1970s) that the theory of quantum mechanics conflicts with classical general relativity and we thus have to find ways to reconcile the two theories. For instance, we try to do so in the form of quantum gravity. However, this is a modern point of view. It was not Einstein’s view. In the 1957 conference at Chapel Hill, Richard Feynman ventured a remark about the possibility of unifying both theories, Quantum mechanics and general relativity (\cite{Kursunoglu}):

\begin{quote}
“Historically, when the unified field theory was first tackled, we had only gravitation, electrodynamics, and a few facts about quantization, so it was natural to try to write down equations which would only unite the Maxwell equations with the gravity equations and leave out of account the strange quantum effects, with the hope that the non-linearity would produce this business. In the meantime, the rest of physics has developed, but still no attempt starts out looking for the quantum effects. There is no clue that a unified field theory will give quantum effects”.    
\end{quote}

\section{Conclusion}

In this comment, I have shown that the ER bridge and the EPR argument were not ”entirely separate” and they were both related to Einstein's work on unified field theory. 

Moreover, I have shown that Brown and Susskind's description uses a modern point of view which is very different from Einstein's own perception and motivation. 
\vspace{5mm} 

This is the third version of my paper which is identical to the second one. But the reason why I am uploading it again is, I want to include Professor Leonard Susskind's interesting and enlightening response to my paper:

\begin{quote}

"Dear Professor Weinstein,

I read your very interesting paper and I don't find anything to disagree with, but I also don't see evidence that E connected entanglement and ERBs in the modern way. 

Einstein himself was an extraordinary phenomenon: I think we will never completely plumb the depths of his intellect.
Did he have any idea that ER=EPR? Who knows? But what is very clear is that the rest of the physics community considered the two ideas to be entirely separate.

Best wishes,

Leonard Susskind"

\end{quote}

Quoted with permission.
\vspace{2mm} 

I am going to think about this response and about the matter further and I hope to write more on the subject in the near future. 

\section*{Acknowledgement}

I would like to thank Scott Aaronson and Gil Kalai for their helpful comments.

\noindent This work is supported by ERC advanced grant number 834735.

\end{document}